\documentclass[twocolumn,showpacs,superscriptaddress,preprintnumbers,amsmath,amssymb,prb]{revtex4-1}
\usepackage{amsmath,amsfonts,bm,graphicx}
\usepackage{color}
\begin{document}

\title{
Inelastic vibrational signals in electron transport across graphene nanoconstrictions
%Inelastic vibrational signals and the role of phonon damping in propensity rules in electron transport across graphene nanoconstrictions\\
%Inelastic vibrational signals in electron transport across graphene nanoconstrictions: The role of phonon damping in propensity rules\\
%Extending propensity rules for inelastic vibrational signals by vibrational damping: Application to electron transport across graphene nanoconstrictions
%electrode modified propensity rules
}

\author{Tue Gunst}
\email{Tue.Gunst@nanotech.dtu.dk}
\affiliation{Department of Micro- and Nanotechnology (DTU Nanotech), Center for Nanostructured Graphene (CNG), Technical University of Denmark, DK-2800 Kgs. Lyngby, Denmark}
\author{Troels Markussen}
\affiliation{QuantumWise A/S, Fruebjergvej 3, Postbox 4, DK-2100 Copenhagen, Denmark}
\author{Kurt Stokbro}
\affiliation{QuantumWise A/S, Fruebjergvej 3, Postbox 4, DK-2100 Copenhagen, Denmark}
\author{Mads Brandbyge}
\affiliation{Department of Micro- and Nanotechnology (DTU Nanotech), Center for Nanostructured Graphene (CNG), Technical University of Denmark, DK-2800 Kgs. Lyngby, Denmark}

\date{\today}

\begin{abstract}
%...\\
%We investigate...
% First-principles - husk i konklussion
We present calculations of the inelastic vibrational signals in the electrical current through a graphene nanoconstriction.
We find that the inelastic signals are only present when the Fermi-level position is tuned to electron transmission resonances, thus, 
providing a fingerprint which can link an electron transmission resonance to originate from the nanoconstriction.
The calculations are based on a novel first-principles method which includes the phonon broadening due to coupling with phonons in the electrodes.
We find that the signals are modified due to the strong coupling to the electrodes, however, still  remain as robust fingerprints of the vibrations in the nanoconstriction.
We investigate the effect of including the full self-consistent potential drop due to finite bias and gate doping on the calculations and find this to be of minor importance.
\end{abstract}

%\pacs{73.22.-f, 72.80.Vp, 73.21.Hb, 73.21.Cd}

\maketitle

\section{Introduction}
Graphene is foreseen to become a versatile material with a wide range of applications in nanoelectronics\cite{bonaccorso_graphene_2015,Novoselov2004,Neto2009}.
Much research addresses the phonon-limited electron mobility of pristine devices\cite{hwang_acoustic_2008,das_sarma_electronic_2011}. However, the properties related to electron-phonon coupling in nanoscale devices based on nanostructured graphene has received much less attention\cite{gunst_phonon_2013,borrnert_lattice_2012}. Nanostructuring of graphene may play a key role in making graphene applicable in electronics since it provides a direct way of tuning the band gap\cite{pedersen_graphene_2008,bai_graphene_2010}, guiding electrons\cite{pedersen_graphene_2012,tombros_quantized_2011,darancet_coherent_2009} as well as tuning the thermal properties.\cite{gunst_thermoelectric_2011,chae_hot_2009,berciaud_electron_2010,jo_low-frequency_2010}
% GNC's
Graphene nanoconstrictions (GNCs) are a generic example of nanostructured graphene that is used for semiconducting interconnects in
graphene nanocircuitry\cite{areshkin_building_2007,botello-mendez_quantum_2011} and may become a central building block of graphene-based nanoelectronics.
State-of-the-art experiments have ``sculpted'' monolayer graphene with close to atomic precision down to a width of a few benzene rings\cite{Xu_controllable_2013,qi_correlating_2014}.
%Advances in bottom-up fabrication can lead to nanoribbons with well-defined edges. 
In addition, a recent experiment indicate how one can control both width and edge morphology of nanoribbons through advances in bottom-up fabrication\cite{ruffieux_-surface_2016}. % of arbitrary nanoribbon structures.
With the emergence of nanosized constrictions the current density can locally be very high and it is important to address the coupling between current and localized vibrations in the device\cite{borrnert_lattice_2012}.

% IETS graphene
% IETS
Recently, several papers have examined inelastic signals due to vibrational excitations in the second derivative of the current-voltage (IV) characteristics, so called Inelastic Electron Transport Spectroscopy (IETS), of gated pristine graphene\cite{zhang_giant_2008,decker_local_2011,brar_observation_2010,Natterer2015,palsgaard_unravelling_2015,lagoute_giant_2015} and heterostructures of graphene and hexagonal boron nitride.\cite{Jung2015,Vdovin2015}
Despite the rapid development in fabrication and electronic characterization\cite{Xu_controllable_2013,qi_correlating_2014} there is to the best of our knowledge still no experimental or theoretical investigation of inelastic vibrational signals for GNCs.

Carbon nanosystems, unlike metallic contacts, have electronic states that vary on the energy scale of the vibrational frequencies necessitating calculations which go beyond the otherwise successful wideband approximated lowest order expansion (LOE-WBA)\cite{frederiksen_inelastic_2004,paulsson_modeling_2005,paulsson_inelastic_2006}. In the LOE-WBA one assumes a constant/energy-independent electronic structure and evaluate all electronic parameters at the Fermi-level. However, phonon frequencies in graphene-based devices can approach $0.2\,$eV on which scale the electronic structure is varying significantly. Hereby it is important to encompass the difference in the electronic states before and after scattering from a vibration.
We have recently developed an extended lowest order expansion (LOE) method that can include the rapid variation near resonances in the electronic spectrum with energy in IETS modeling\cite{lu_efficient_2014}. This method enables studies of IETS on gated graphene nanostructures.
IETS was originally developed to probe molecules on surfaces with scanning tunneling microscopy (STM) that are weakly bound to the leads\cite{galperin_molecular_2007} therefore possessing a set of localized vibrations. In the case of nanostructured graphene the vibrations of the device is strongly coupled with phonons in both leads and the resulting life-time broadening needs to be included in a predictive description of the inelastic transport signals\cite{gagliardi_electronphonon_2008,engelund_atomistic_2009}. The so-called \textit{propensity rules}, approximate selection rules related to the symmetry of vibrational modes and electronic states of the junction, explain why only a few of many possible vibrational modes yield an inelastic signal\cite{paulsson_unified_2008,kim_scattering_2013,gagliardi_textitpriori_2007,troisi_propensity_2006}. The life-time broadening could be severe and therefore needs to be considered in strong-coupled devices.

In this paper, we apply the extended LOE method\cite{lu_efficient_2014} to describe the inelastic vibrational signals in the current for a GNC near an electronic resonance including the phonon damping from the leads.
The simulations are performed with DFT and nonequilibrium Green's functions (DFT-NEGF) packages\cite{soler_siesta_2002,brandbyge_density-functional_2002,frederiksen_inelastic_2007,FN_settings} in combination with Inelastica\cite{frederiksen_inelastic_2007}. In addition, we have implemented both the LOE-WBA and LOE methods in the Atomistix ToolKit (ATK) simulation tool\cite{ATK} to be able to compare the two methods. We find consistent results with both Inelastica and ATK which strengthens the reproducibility of the results. We identify a number of inelastic vibrational signals in the current which persists including the strong coupling to electrode phonons in the GNC. We furthermore determine the impact on the IETS signals of finite bias and charge doping due to gate electrodes in the self-consistent calculation of the Hamiltonian.

%Phonon broadening\cite{gagliardi_electronphonon_2008}
%IETS\cite{lu_efficient_2014}
\begin{figure}[!htbp]%[!t]%[!htbp]%
\centering
{\tiny }{\includegraphics[width=0.99\linewidth]{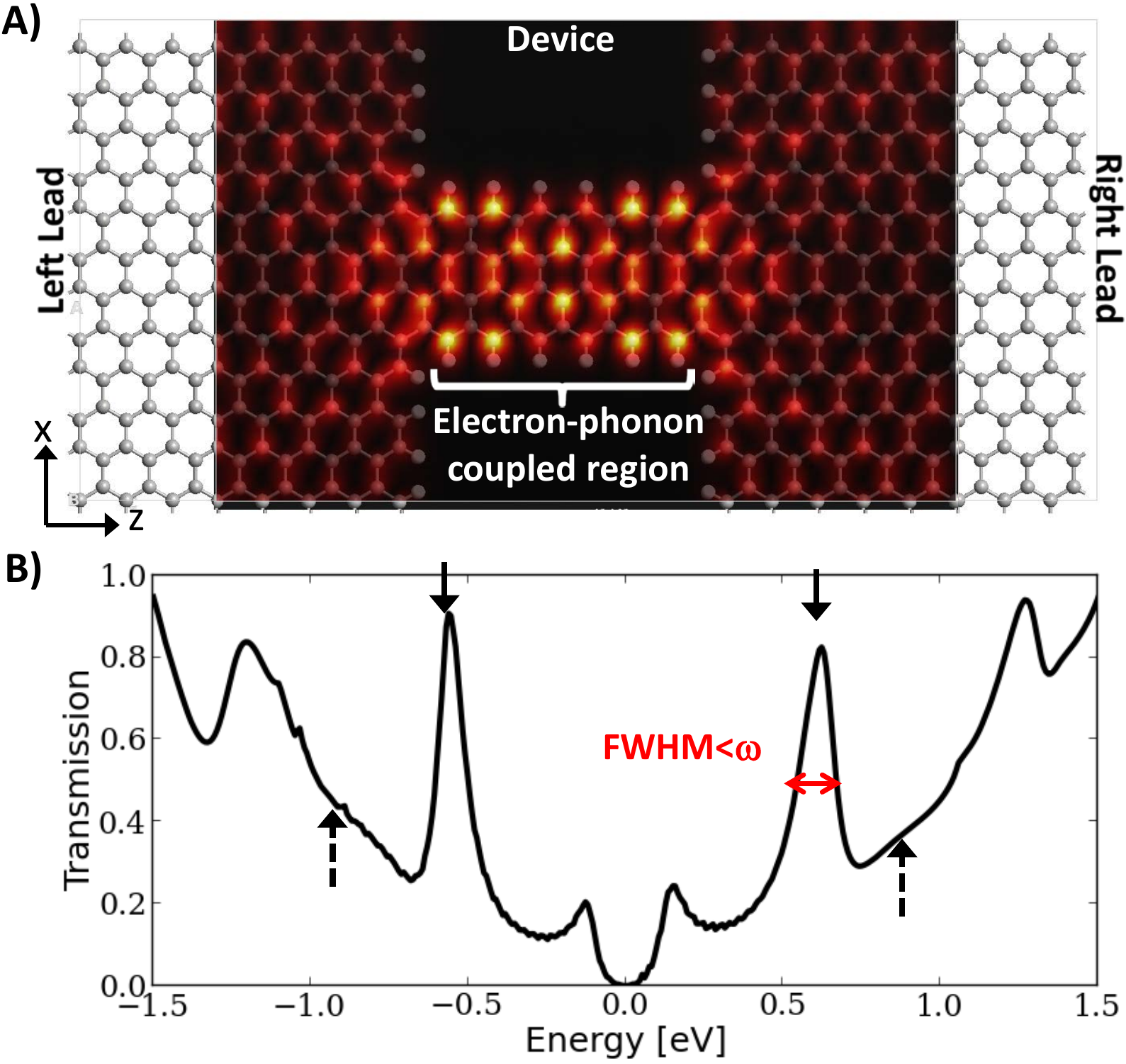}}\\%Exact energies: 0.62 and -0.56
{\includegraphics[width=0.99\linewidth]{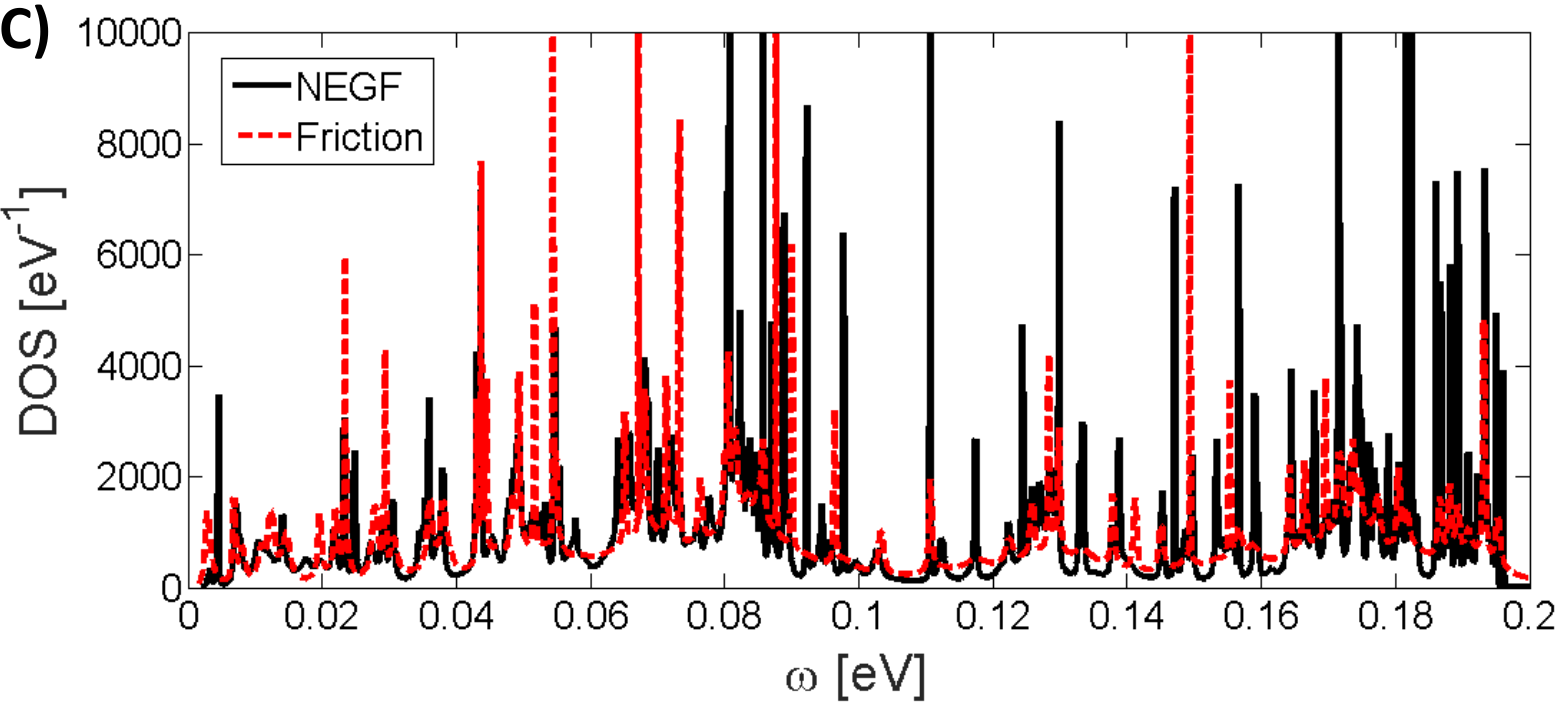}}
\caption{(Color online) System and transmission for a nanostructured graphene device. A) Graphene nanoconstriction with a high current density  at the narrow ribbon connecting two graphene electrodes. The transmission eigenchannel at the $\mu_F\approx$0.6eV is plotted on top of the configuration indicating the path of the current. The electron-phonon coupling is evaluated within the constriction where the current density is large. B) Transmission across the device showing several resonances with a full width at half maximum (FWHW) lower than typical phonon frequencies.  On (Off) resonance positions are indicated by full (dashed) arrows. C) Phonon density of states comparing the result from full NEGF with the friction approximation.}
\label{fig:SystemAndTrans}
\end{figure}
\section{System and method}
We consider the GNC system illustrated in Fig.~\ref{fig:SystemAndTrans}A where the current is passed through a short ribbon\cite{Brey2006,christensen_identification_2015,ruffieux_-surface_2016} at the narrowest point that connects two graphene electrodes.
Looking at the transmission probability for an electron to cross the device, Fig.~\ref{fig:SystemAndTrans}B, we find that several electronic resonances are present due to the diffraction barrier at abrupt interfaces in graphene\cite{darancet_coherent_2009,ihnatsenka_conductance_2012}.
This diffraction barrier height is controlled by the width of the constriction. Making the constriction longer will move the peaks down in energy while maintaining the overall features. Here we choose a length where inelastic calculations are computationally heavy but still feasible. 
A gate-electrode can be used to control the Fermi-level and electronic states involved in the transport.
Clearly the transmission probability, Fig.~\ref{fig:SystemAndTrans}B, varies significantly on the scale of typical optical phonon frequencies ($0.2\,$eV).
%gating: Gating the device will enable one to probe electron transport near the different resonances.

% \section{Method}
%We start by briefly reca
The basis of the LOE method is the Meir-Wingreen formula for the electron current where one in addition apply a set of closed Dyson and Keldysh equations by replacing the
full Green's function with the single-particle Green's function\cite{lu_efficient_2014}.
The equations are expanded to lowest order in the electron-phonon coupling matrix ($\mathbf{M}^{\lambda}$) in the device region and simplified to describe the IETS signals using the fact that these are prominent only close to the excitation threshold where the applied bias equals the vibrational energy, 
$V_b=\mu_L-\mu_R=\pm \omega_\lambda$. Here $\mu_{L/R}$ are the chemical potentials of the left/right electrodes, and $\omega_\lambda$ the vibrational energy (we employ atomic units unless explicitly stated, $e=\hbar=1$). The LOE expression for the second derivative of the current, $I$,  for a given mode, $\lambda$,  is a sum of two analytical functions\cite{lu_efficient_2014},

\begin{equation}
\label{eq:current}
\partial_{V}^2{I}=  \gamma_\lambda \, \partial_{V}^2{\cal I}^\mathrm{sym}(V,\omega_\lambda, T) +  \kappa_\lambda\,\partial_{V}^2{\cal I}^\mathrm{asym}(V,\omega_\lambda, T)
\end{equation}
where,
\begin{eqnarray}
	\mathcal{I}^\mathrm{sym}\! &\equiv&\! \frac{{G}_0}{2}\sum_{\sigma=\pm} \!\!\!\sigma(\omega_\lambda+\sigma V)\\
&&\times\!\left(\!\coth\!\frac{\omega_ \lambda}{2k_BT}-\coth\!\frac{\omega_\lambda+\sigma V}{2k_BT}\!\right)\nonumber.
\end{eqnarray}
and
\begin{eqnarray}
\mathcal{I}^\mathrm{asym} &\equiv& \frac{{G}_0}{2}\int_{-\infty}^{+\infty}\!d\varepsilon\mathcal{H}\{ f(\varepsilon'_-)- f(\varepsilon'_+)\}(\varepsilon)
\\
&&\times\left(f(\varepsilon-eV)-f_{}(\varepsilon)\right) \nonumber\,,
\end{eqnarray}
where $f$ is the Fermi-Dirac function, $\varepsilon_{\pm}'=\varepsilon \pm  \omega$ and $G_0$ the conductance quantum.
The prefactors can be expressed in terms of the unperturbed retarded/advanced Green's function $\mathbf{G}^{r/a}$, and the (time-reversed) spectral density matrices $\mathbf{A}_{\alpha}=\mathbf{G}^{r}\mathbf{\Gamma}_{\alpha}\mathbf{G}^{a}$ ($\tilde{\mathbf{A}}_{\alpha}=\mathbf{G}^{a}\mathbf{\Gamma}_{\alpha}\mathbf{G}^{r}$), and only involve evaluations of these quantities at the chemical potentials for the corresponding excitation threshold ($\mu_L-\mu_R=\pm\omega_\lambda$). We have
$\gamma_\lambda=\gamma_{i,\lambda}+\gamma_{e,\lambda}$, with $\gamma_{e,\lambda}\approx{\rm Im} B_\lambda$, $\kappa_{\lambda}=2{\rm Re} B_\lambda$, 
\begin{eqnarray}
\label{eq:ic4af}
\gamma_{i,\lambda}&=&{\rm Tr}\!\left[ \mathbf{M}^{\lambda} \tilde{\mathbf A}_L(\mu_L) \mathbf{M}^{\lambda} \mathbf A_R(\mu_R) \right], \label{eq:gamma-i}
\end{eqnarray}
and
%\begin{widetext}
\begin{eqnarray}
B_\lambda &\equiv&{\rm Tr}[\mathbf{M}^{\lambda} \mathbf A_R(\mu_L)\mathbf\Gamma_L(\mu_L)\mathbf G^r(\mu_L)\mathbf{M}^{\lambda}\mathbf A_R(\mu_R)\nonumber\\
&-&\mathbf{M}^{\lambda} \mathbf G^a(\mu_R)\mathbf\Gamma_L(\mu_R)\mathbf A_R(\mu_R)\mathbf{M}^{\lambda} \mathbf A_L(\mu_L)]. \label{eq:ic4ag}
\end{eqnarray}
The first part, Eq.~\eqref{eq:gamma-i}, is related to the Fermi's Golden rule rate of scattering from an incoming state with energy  $\varepsilon$, to a final state with energy $\varepsilon_{\pm}'$.
However, the dependence on energy is more complicated for the remaining interference terms, Eq.~\eqref{eq:ic4ag}.

These results are based on a non-interacting (infinite life-time) phonon spectral density given by,
\begin{eqnarray}
%\mathcal{A}(\omega)=2\pi\sum_{\lambda}\left(\delta(\omega-\omega_{\lambda})-\delta(\omega+\omega_{\lambda})\right)\label{eq:ph_spectral_density}\,.
\mathcal{A}(\omega)=2\pi\sum_{\lambda}\left(\mathcal{L}(\omega-\omega_{\lambda})-\mathcal{L}(\omega+\omega_{\lambda})\right)\label{eq:ph_spectral_density}\,,
\end{eqnarray}
with $\mathcal{L}(\omega)=\delta(\omega)$ for zero temperature and zero coupling to the electrodes.
A broadening from the electrode phonons can be included as a post-processing for each mode by convoluting the $\partial_{V}^2{I}(V)$ signal with the device vibrational spectral function including the coupling to the electrode phonons. This can be seen using the Lehmann representation, see Viljas {\it et al.}\cite{ViCuPa.05.Electron-vibrationinteractionin}. In the simplest case we may use a broadened delta-function%, generalizing the delta function at the threshold voltage in Eq.~\eqref{eq:ph_spectral_density}.
\begin{eqnarray}
\mathcal{L}(\omega=V_{SD})=\frac{1}{\pi} \frac{\eta_{ph}/2}{(\eta_{ph}/2)^2 + V_{SD}^2} \,, \label{eqn:broadening}
\end{eqnarray}
to obtain the signal at threshold voltage as
\begin{eqnarray}
\label{eq:currentBroadened}
\partial_{V}^2{I}_B(V)&=&  \int dV' \partial_{V'}^2{I}(V') \mathcal{L}(V-V')\,.
\end{eqnarray}
The broadening, or linear friction coefficient, can be calculated from the diagonal elements of the phonon self-energy due to the coupling with the leads, $\eta_{ph}=-\frac{\partial}{\partial \omega}\left(\rm{Im}[\Pi_{ph}^r]\right)|_{\omega=0}$. Alternatively, we may use the actual phonon density of states (DOS) of each mode:
\begin{eqnarray}
\mathcal{L}(\omega)= \rm{DOS}(\lambda,\omega) = - \frac{2 \omega}{\pi} \rm{Im}[\mathbf{D}^r_{\lambda,\lambda}(\omega)] \,, \label{eqn:broadening2}
\end{eqnarray}
where we made use of the phonon retarded Green's function $\mathbf{D}^r$ expressed in the phonon mode eigenspace. Both neglects coupling between vibrations mediated by the electrode phonons.
The broadening in Eq.~\eqref{eqn:broadening2} has a more complex lineshape than the Lorentzian, but is guaranteed to reproduce all features in the total phonon $\rm{DOS}=\sum_{\lambda}\rm{DOS}(\lambda,\omega)$.
%phonon spectrum such as the $\mathrm{Tr}[]$

In Fig.~\ref{fig:SystemAndTrans}C, we compare the DOS found from NEGF, cf. Eq.~\eqref{eqn:broadening2}, with that of the friction model, using the approximate self-energy $\Pi_{ph}^r=-i\eta_{ph}\omega$ in accordance with the Lorentzian broadening in Eq.~\eqref{eqn:broadening}, to substantiate our broadening models.
The friction model is able to capture most key signatures in the DOS\cite{Note1}.
%On the other hand it is guaranteed to underestimate the broadening consistently.

\section{Inelastic simulations and results}
\begin{figure}[!htbp]%[!t]%[!htbp]%
\centering
{\includegraphics[width=0.99\linewidth]{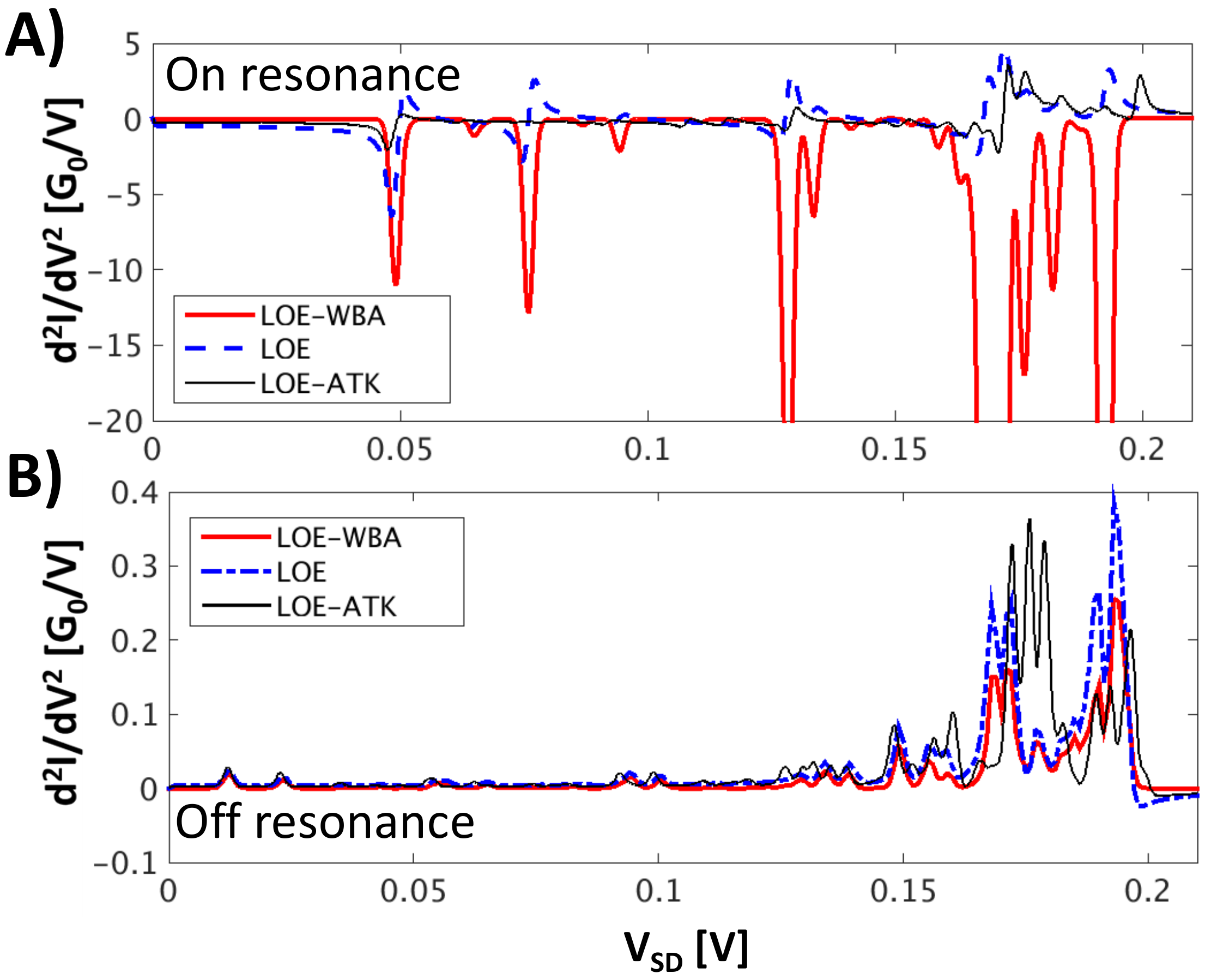}}\\
{\includegraphics[width=0.99\linewidth]{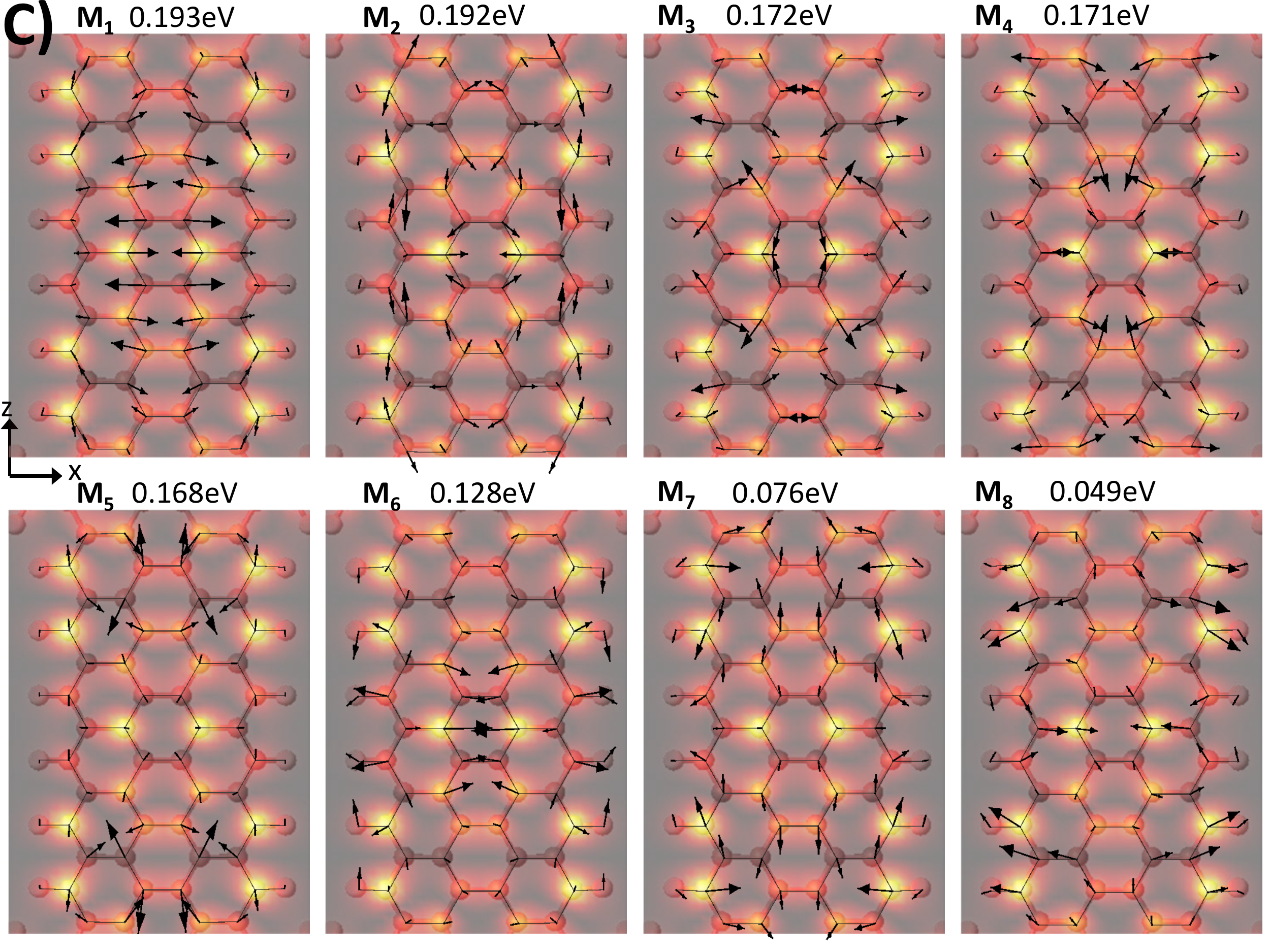}}
\caption{(Color online) Comparison of the LOE-WBA (full red line) and LOE (dashed blue line) results for the inelastic vibrational signals in the second derivative of the current. The device is either gated to A) a transmission resonance at $\mu_F\approx \pm$0.6eV or B) away from resonance at $\mu_F\approx \pm$0.85eV. We also compare with the LOE-results obtained with ATK (thin black line). C) Vibrational modes contributing to the five main peaks in the second derivative of the current at resonance. The mode displacements are illustrated by arrows, on top of the current density from Fig.~\ref{fig:SystemAndTrans}A, inside the constriction. The transport direction is along $z$ (opposite to Fig.~\ref{fig:SystemAndTrans}A).}
\label{fig:IETSandModes}
\end{figure}

%We start by analyzing the results that do not include the effect from the finite bias electronic structure and an actual gate electrode.
We now apply the widely used model gate where one simply tune the Fermi-level, $\mu_F$, through the electronic spectrum. We start by analyzing results neglecting the phonon broadening from the electrodes.
In Fig.~\ref{fig:IETSandModes} the results from the LOE are presented and compared to that of the original LOE-WBA where all electronic parameters are
evaluated at the equilibrium chemical potential.
We compare two situations: one where the system is gated close to the resonance ($\mu_F\approx \pm$ 0.6eV), Fig.~\ref{fig:IETSandModes}A, or one where the system is gated to a chemical potential where the electronic structure is effectively energy independent within the scale of phonon frequencies ($\mu_F\approx \pm$ 0.85eV), Fig.~\ref{fig:IETSandModes}B. On and off resonance positions are also indicated by arrows in Fig.~\ref{fig:SystemAndTrans}B.
At resonance, see Fig.~\ref{fig:IETSandModes}A, the spectrum changes quite remarkably between the two models. The LOE method gives rise to several dip-peak features not present in the original LOE-WBA model.
Within the LOE-WBA the electron-phonon coupling seems artificially strong, i.e. the change of conductance is larger than a few percents.
The LOE gives a significantly lower signal which is related to the difference in density of states for the initial and final states.
Tuning the energy away from the resonance ($\mu_F\approx \pm$ 0.85eV), the LOE model gives results consistent with the LOE-WBA, Fig.~\ref{fig:IETSandModes}B.
The IETS signal undergoes a sign change from peaks off resonance to dips at resonance, consistent with previous single-level model considerations\cite{paulsson_unified_2008,kim_scattering_2013}. In addition, we have shown the results obtained from the LOE implemented in ATK Fig.~\ref{fig:IETSandModes}A,B. In general, we find good agreement between the two implementations. The same signals are present in both calculations with differences in peak position and intensity being related to a slight variation in the equilibrium lattice constant in the two DFT codes using different pseudopotentials.
%\textcolor[rgb]{1,0,0}{Should we explain why peak off resonance ($T<0.5$) and dips at resonance ($T>0.5$)?}

We will next analyze the origin of the five distinct peaks at resonance where the strongest interacting modes are located around $170\,$meV.
The eight phonon modes with the largest inelastic signal are illustrated in Fig.~\ref{fig:IETSandModes}C. The contributing modes are the same within both LOE-WBA and LOE although the LOE signal is clearly different. 
Since the current mainly runs through the $\pi$-orbitals, we expect the current to interact the strongest with longitudinal modes in the device plane. Due to the symmetry plane the Hamiltonian is the same for planar graphene nanostructures whether we move atoms up or down in the out-of-plane direction. Therefore, the out-of-plane electron-phonon coupling elements between $\pi$-orbitals will be zero and the characteristic vibrations found for the GNC are all in-plane modes as expected.
Comparing the scattering state symmetry at resonance, Fig.~\ref{fig:SystemAndTrans}A and repeated inside the constriction in Fig.~\ref{fig:IETSandModes}C, it is evident that these modes all have displacement in the regions where the scattering state and current density is largest, i.e. near the ribbon edge of the entrance to the constriction or near the center of the ribbon.

With an explanation of the vibrational signals near and far from resonance at hand, we now target three additional questions. Firstly, we apply the extensively used approximation of a rigid shift as a gate voltage to screen the IETS on a fine grid of gate ($V_g$) and source-drain bias ($V_{SD}$) voltages. This enable us to evaluate how close to the resonance we need to gate before strong IETS appear. Secondly, we will apply the broadening from the electrode phonons in order to evaluate the robustness of the signals. Finally, we will include the self-consistent electronic structure obtained at a finite bias and gate doping which is a quite demanding calculation, but enable us to judge the importance of including the full self-consistent potential, which was so far neglected.

\begin{figure}[!htbp]%[!t]%[!htbp]%
\centering
{\includegraphics[width=0.99\linewidth]{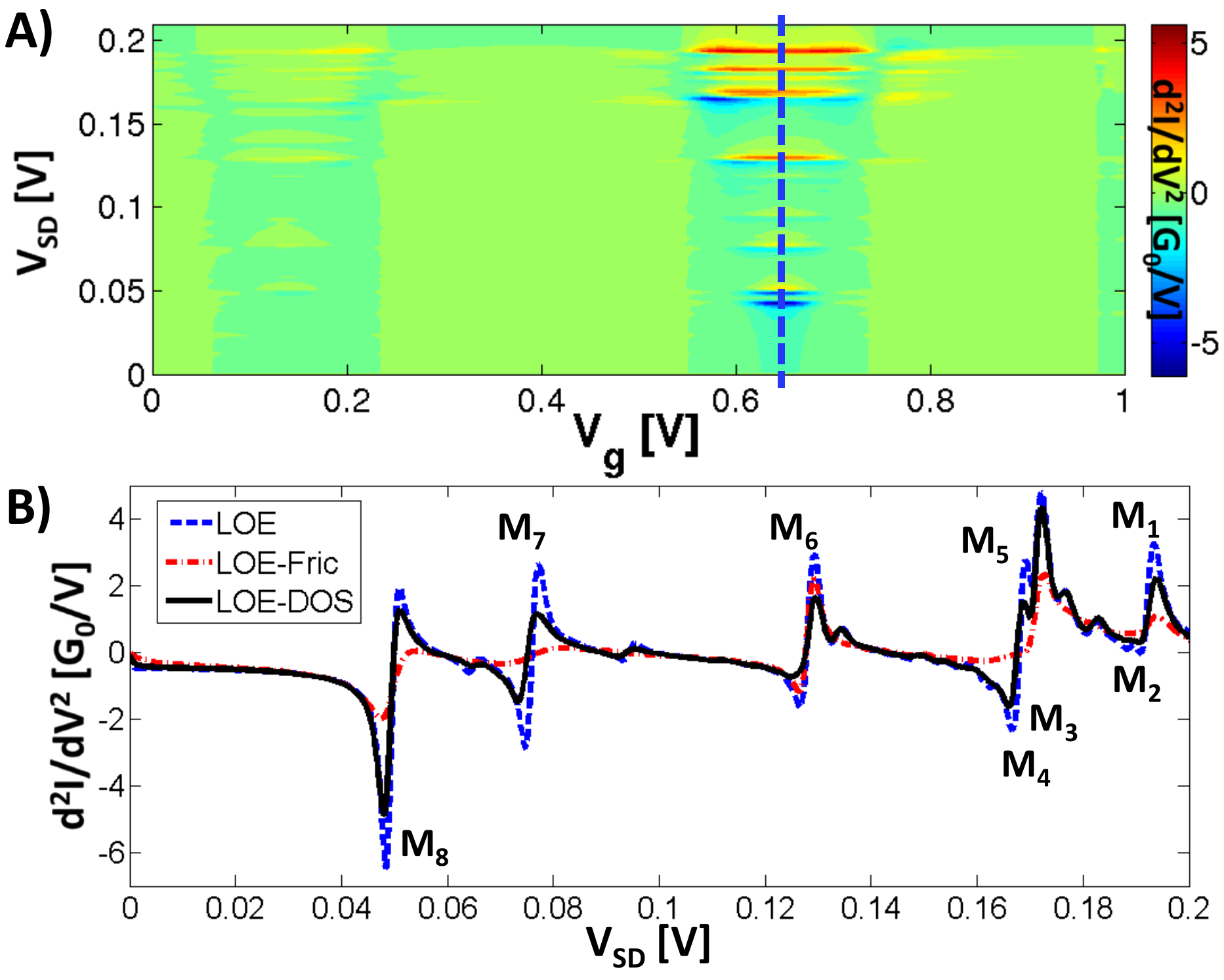}}
%\\{\includegraphics[width=0.99\linewidth]{Figures/Fig3C_Zoom.pdf}}
\caption{(Color online) A) Inelastic vibrational signals in the LOE second derivative of the current as a function of gate, $V_g$, and applied bias voltages, $V_{SD}$. B) Second derivative of the current at the specific gate value indicated by a dashed vertical line (on resonance) in A).
%C) Zoom into the boxed region indicated in B) showing how one signal almost vanishes while another survives the phonon broadening.
}
\label{fig:GateBiasAndBroadening}%Dashed vertical line indicates the gate value for the plot in B).
\end{figure}

The computed IETS signals as a function of varying gate voltage are shown in Fig.~\ref{fig:GateBiasAndBroadening}A as a density plot. 
It illustrates how the IETS signal is largest for gate values where the transmission varies the most, i.e. at the resonance ($V_g\approx$0.6eV, vertical dashed line) and at the band edge ($V_g\approx$0.15eV). In addition, the signals are clearly present in a region of $\Delta V_g\approx  \omega$ (up to 0.2\,V) around the peak position. We conclude that the inelastic signals are only present when the Fermi-level position is tuned to gate values where electron transmission resonances are present. Therefore IETS, and its variation with gate voltage, will strongly indicate if sharp resonances are present in a nanostructured graphene device.
The IETS spectra at $V_g\approx$0.6eV is shown in Fig.~\ref{fig:GateBiasAndBroadening}B.
So far the finite broadening in the vibrational signals was obtained from the finite temperature of $T=4.2$K. Other broadening mechanisms exist, e.g. originating from a lock-in modulation voltage\cite{paulsson_inelastic_2006} or coupling to the surrounding electrode phonon baths, and anharmonic couplings.
% FWHM = 5.4 kBT = 2.0 meV, FWHM = 1.7 Vrms = 13.6 meV for Vrms = 8 meV.
%\begin{center}
\begin{table}[!htbp] %htb
%\begin{ruledtabular}
\begin{tabular}{|c|c|c|c|c|c|c|c|c|}
\multicolumn{9}{c}
	%{\rule[-3mm]{0mm}{8mm}\textbf{Table of damping}} \\
	{\rule[0mm]{0mm}{0mm}}\\
  \hline
  & \textbf{M1} & \textbf{M2} & \textbf{M3} & \textbf{M4} & \textbf{M5} & \textbf{M6} & \textbf{M7} & \textbf{M8} \\\hline
  $\omega$ [meV] & 193&  192 & 172 & 171 & 168 & 128 & 76 & 49 \\\hline
  $\eta_{ph}$ [meV] & 16.2 & 3.0 & 0.6 & 5.2 & 6.3 & 0.2 & 6.4 & 2.6 \\\hline
  %$\eta_{e}$ [meV]  & 12.9 & 8.0 & 15.1 & 7.9 & 26.0 & 5.8 & 1.8 & 1.0 \\
  $\eta_{e}$ [meV]  & 0.9 & 0.6 & 1.7 & 0.9 & 2.9 & 0.2 & 1.9 & 1.1 \\
  \hline
\end{tabular}\caption{The friction from phonon, $\eta_{ph}$, and electron, $\eta_{e}$, baths for the eight characteristic modes.}\label{Tbl:Damping}
%\end{ruledtabular}
\end{table}
%\end{center}
Table~\ref{Tbl:Damping} contains the linear frictions, giving the broadening due to the phonon bath, for the characteristic modes. The friction is smallest for modes localized in the center of the constriction (e.g. M3 and M6) and larger for modes with displacements near the contacts (e.g. M1, M5 and M7). The largest phonon friction in the system is found to be $\eta_{ph}=77\,$meV for comparison for a mode that however does not couple significantly with the current.
We find that the phonon broadening vary by three orders of magnitude, in the range $0.1-100\,$meV, between the different phonon modes. 

In Fig.~\ref{fig:GateBiasAndBroadening}B, we include the damping/broadening from the friction model, red dashed-dotted line obtained from Eq.~\eqref{eqn:broadening}, and the full phonon DOS, black solid line obtained from Eq.~\eqref{eqn:broadening2}, and compare with the original signal, blue dashed line.
We include the broadening from the electrode phonons as the convolution described in Eq.~\eqref{eq:currentBroadened}.
Despite the broadening from electrode phonons we find robust fingerprint signatures.
In addition, the difference in the broadening between the vibrational modes is clearly visible in the IETS signals. For instance, the signal around 0.08 and 0.2\,eV (M1, M2, M7) is reduced significantly while most of the signals at 0.05, 0.128 and 0.17\,eV (M8, M6, M3-5) survives.
The frictional broadening model, cf. Eq.~\eqref{eqn:broadening}, exaggerates the broadening mechanism compared to the full lineshape model, cf. Eq.~\eqref{eqn:broadening}. However, M6 is reduced slightly more by the full lineshape model than the friction model.
%An example of the difference in the signal broadening is highlighted in Fig.~\ref{fig:GateBiasAndBroadening}C (M7, M8).

The line-shape itself can change due to the phonon broadening such that a dip-peak resemble more a peak when the phonon broadening is included, see for instance the highest frequency mode (M1). We conclude that the dominant inelastic vibrational signals occur for modes that at one time has a symmetry dictated by the electronic scattering states in Fig.~\ref{fig:IETSandModes}A and at the same time is marginally localized near the electrodes so that the vibrational broadening from the electrode phonons is low (Fig.~\ref{fig:IETSandModes}C and Table~\ref{Tbl:Damping}).\\
We have also tried to apply a constant artificial broadening to all peaks, to examine at what friction-value the peaks start to vanish. We find that the first peaks get impossible to distinguish at a friction of approximately $5\,$meV while all peaks vanish at constant frictions above $10\,$meV. Both of these values are smaller than typical broadenings found in the system.

In Table~\ref{Tbl:Damping} we for comparison list the damping due to electronic friction, $\eta_{e}$, calculated from the method described in Ref.~\onlinecite{gunst_phonon_2013}.
%$=\frac{1}{2\pi} \sum_{\alpha,\beta}{\rm{Tr}}[\mathbf{M}^{\lambda}{\bf A}_\alpha(\mu_\alpha)\mathbf{M}^{\lambda}{\bf A}_\beta(\mu_\beta)]$.
The electronic friction is in general strongly dependent on bias voltage and is here evaluated at the threshold voltage $V_{SD}=\omega_{\lambda}$. It is notable large for mode M5 and can for a few modes (M3, M6) be on the same order of magnitude as the phonon friction. A few modes with a strong coupling to the current also obtain a large phonon friction relative to the electronic friction, i.e. M1 and M7. Interestingly, we find that $\eta_{e}$ decreases with $V_{SD}$ since the electronic structure away from the resonance comes into play. As a consequence the electronic friction may play a more dominant role as broadening near a resonance, while it can be tuned with the applied bias voltage. In a previous study we calculated the current-induced forces and heating in the GNC system. We note that the modes here giving the largest IETS signal in the current are different from the modes which we have found to yield a highly nonlinear heating at bias voltages above 0.4\,V, and which are related to a current-induced and can give rise to negative electronic friction for certain "run-away" modes\cite{gunst_phonon_2013}.

\begin{figure}[!htbp]%[!t]%[!htbp]%
\centering
{\includegraphics[width=0.99\linewidth]{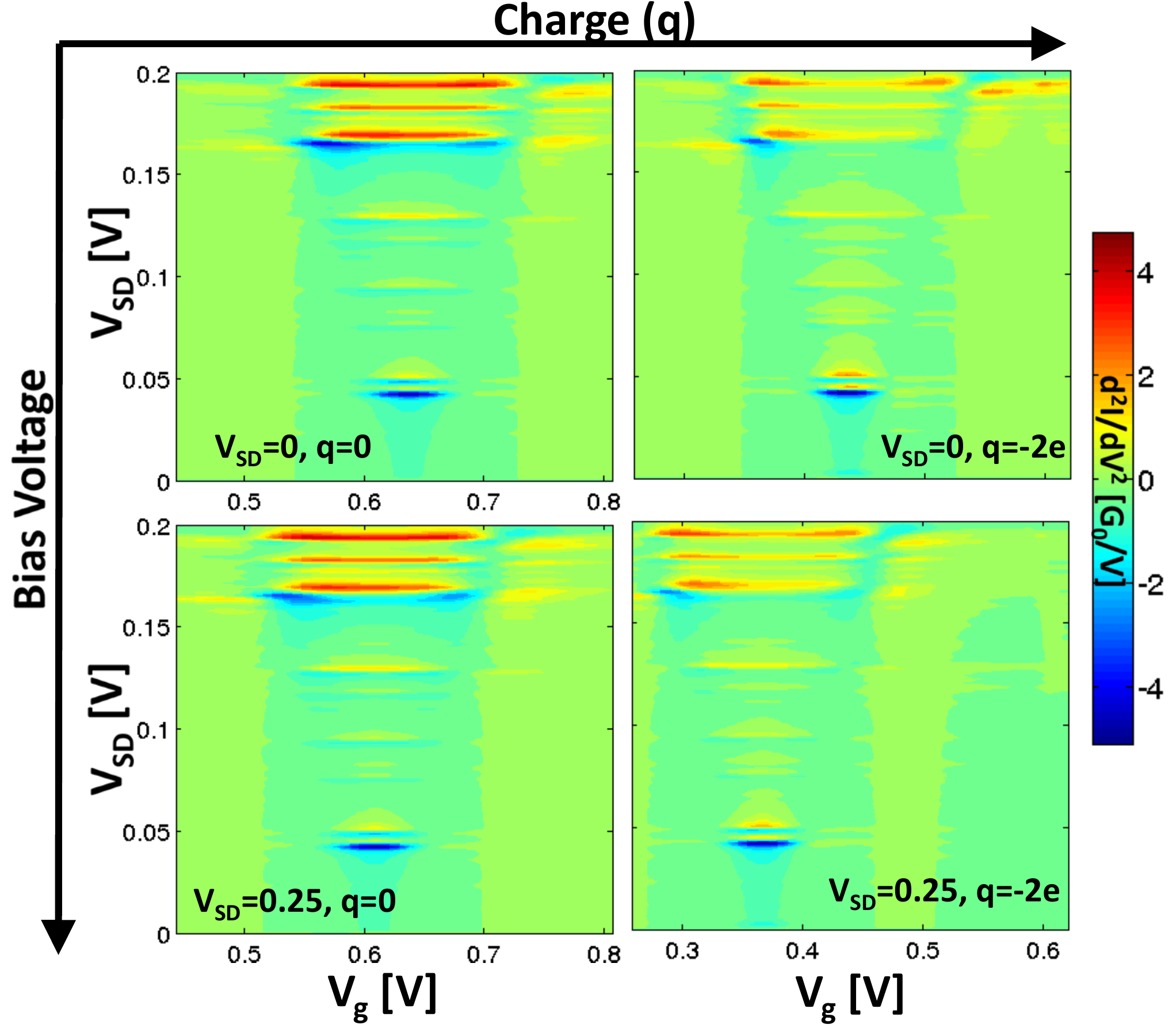}}
\caption{(Color online) Dependence of the second derivative of the current on the physical gate and bias voltage.}
\label{fig:BiasAndChargeIETS}%Dashed vertical line indicates the gate value for the plot in B).
\end{figure}
In Fig.~\ref{fig:BiasAndChargeIETS} we include the self-consistent electronic structure obtained at a finite bias and an electrostatic gate. The electrostatic gate is modeled by adding a charge, $q$, in a plane $20\,${\AA} below the system and $-q$ to the device generating an electric field. For a detailed discussion of the gating model and the potential drop we refer the reader to Ref.~\onlinecite{papior_manipulating_2015}. Here, we focus on the main signal near resonance. Increasing the bias voltage in the DFT-NEGF simulation ($V_{SD}=0.25$\,V and $q=0$), and hereby including the potential drop in the electronic structure, introduces a slight shift in the signal position with respect to $V_g$, but does not influence its magnitude. Changing the charge of the system separately ($V_{SD}=0$\,V and $q=-2e$) is observed to move the location of the signal almost 0.2\,eV.
Including both charge and bias voltage ($V_{SD}=0.25$\,V and $q=-2e$) does to some extend break the symmetry of the IETS signals but the magnitude and the dominating modes are unaffected.

\section{Conclusions}
% Short version:
In summary, we have presented the first calculations of inelastic vibrational signals in GNCs, where the phonon broadening and high phonon frequencies necessitates extended methodology. First-principles calculations of the inelastic vibrational signals in the current can include the electrode phonon broadening through a post-processing for each vibrational mode. In particular, we find that the broadening from the electrode phonons can vary by up to three orders of magnitude between the vibrational modes. 
We find several strong inelastic signals for Fermi-level positions close to electron transmission resonances which are robust against finite-bias effects as well as broadening from the electrode phonons. Therefore, inelastic signals depending on gate voltage can be used to investigate if sharp electron transmission resonances are present in a nanostructured graphene device.
The propensity rules dictate that the dominant inelastic vibrational signals occur for modes that both has a symmetry coinciding with that of the electronic scattering states and at the same time is marginally localized near the electrodes so that the vibrational broadening from the electrode phonons is low.

\section{acknowledgement}
% vibrational fingerprints
%We thank J.-T. {L\"u} (Wuhan) and T. Frederiksen (San Sebastian) for useful discussions on the methods employed.
The authors acknowledges support from Innovation Fund Denmark, grant Nano-Scale Design Tools for the Semiconductor Industry (j.nr. 79-2013-1).
The Center for Nanostructured Graphene (CNG) is sponsored by the Danish Research Foundation, Project DNRF103.

%\nocite{*}
%\bibliographystyle{unsrt}

%

\end{thebibliography}

%\bibliography{additionalRefs,BibTexs/BibOther2DMaterials,BibTexs/Graphene,BibTexs/BibDiverse,BibTexs/BibGrapheneElectron-phononcoupling,BibTexs/BibElectron-Phononcoupling,BibTexs/books,BibTexs/BibBoltzmannModels,BibTexs/BibSiNWs,BibTexs/BibBoltzmannTransportEquationAndMethods,BibTexs/BibOwnPapers,BibTexs/BibCID_Theory,BibTexs/BibDFTandSiestaMethods,BibTexs/BibIETS,BibTexs/BibGrapheneConstrictionsAndRibbons,BibTexs/BibThermoelectrics,BibTexs/BibHighBiasAndMechanics,BibTexs/BibGrapheneOverview,BibTexs/BibGrapheneConstrictionAndRibbonsInelastic,BibTexs/BibGrapheneAntidotlattices,BibTexs/BibGrapheneWaveguides,BibTexs/BibFabrication,BibTexs/BibFootnotes}

%\bibliography{BibTexs/BibFootnotes,BibTexs/BibIETS,BibTexs/BibGrapheneConstrictionsAndRibbons,BibTexs/BibGrapheneConstrictionAndRibbonsInelastic,BibTexs/Graphene,BibTexs/BibDFTandSiestaMethods,BibTexs/BibGrapheneWaveguides,BibTexs/BibHighBiasAndMechanics,BibTexs/BibThermoelectrics,BibTexs/BibGrapheneOverview,BibTexs/BibIsotopeEffect,BibTexs/BibGrapheneAntidotlattices,BibTexs/BibHeatTransport,BibTexs/BibFabrication,BibTexs/BibRamanGraphene,BibTexs/BibGrapheneElectronTransport,BibTexs/BibDiverse,BibTexs/BibElectronicStates,BibTexs/BibMagneticProperties,BibTexs/BibCarbonNanotubes,BibTexs/BibOwnPapers,BibTexs/books,BibTexs/NEGFthermoelectricEph,BibTexs/BibAdditionalThesisRefs,BibTexs/BibQMDsimulations,BibTexs/BibCID_Theory,BibTexs/BibNEGFtheory,BibTexs/BibThermalExpansion,BibTexs/BibGrapheneBendingRipplingAndLineDefects,BibTexs/BibSubstrate,BibTexs/BibTopologyOptimization}

\begin{thebibliography}{1}
\bibitem [{FN_()}]{FN_settings}%
  \BibitemOpen
  \href@noop {} {\ }\bibinfo {note} {We used single-zeta-polarized (SZP)
  localized pseudoatomic orbitals in the basis set for both hydrogen and
  carbon. Exchange-correlation effects were handled within GGA-PBE and a grid
  cutoff of 210 Ry was applied. The Brillouin zone was sampled with a [5 1 2]
  Monkhorst-Pack grid, with the x-axis as the transverse direction and the
  z-axis as the transport direction. The system was relaxed until all forces
  were smaller than 0.005eV/Ang. We used 150 points along the bias contour with
  a broadening of 0.01 eV.}\BibitemShut {Stop}%
\end{thebibliography}

\end{document}